\documentclass[12pt]{article}
\usepackage{graphicx}
\begin{document}
\newcommand{\beq}{\begin{equation}}
\newcommand{\eeq}{\end{equation}}
\newcommand{\beqa}{\begin{eqnarray}}
\newcommand{\eeqa}{\end{eqnarray}}
\newcommand{\beqar}{\begin{eqnarray*}}
\newcommand{\eeqar}{\end{eqnarray*}}
\newcommand{\al}{\alpha}
\newcommand{\be}{\beta}
\newcommand{\del}{\delta}
\newcommand{\D}{\Delta}
\newcommand{\eps}{\epsilon}
\newcommand{\ga}{\gamma}
\newcommand{\Ga}{\Gamma}
\newcommand{\ka}{\kappa}
\newcommand{\nn}{\nonumber}
\newcommand{\inn}{\!\cdot\!}
\newcommand{\h}{\eta}
\newcommand{\ii}{\iota}
\newcommand{\kk}{\varphi}
\newcommand\F{{}_3F_2}
\newcommand{\la}{\lambda}
\newcommand{\La}{\Lambda}
\newcommand{\na}{\prt}
\newcommand{\Om}{\Omega}
\newcommand{\om}{\omega}
\newcommand{\p}{\phi}
\newcommand{\sig}{\sigma}
\renewcommand{\t}{\theta}
\newcommand{\z}{\zeta}
\newcommand{\ssc}{\scriptscriptstyle}
\newcommand{\eg}{{\it e.g.,}\ }
\newcommand{\ie}{{\it i.e.,}\ }
\newcommand{\labell}[1]{\label{#1}} 
\newcommand{\reef}[1]{(\ref{#1})}
\newcommand\prt{\partial}
\newcommand\veps{\varepsilon}
\newcommand{\pol}{\varepsilon}
\newcommand\vp{\varphi}
\newcommand\ls{\ell_s}
\newcommand\cF{{\cal F}}
\newcommand\cA{{\cal A}}
\newcommand\cS{{\cal S}}
\newcommand\cT{{\cal T}}
\newcommand\cV{{\cal V}}
\newcommand\cL{{\cal L}}
\newcommand\cM{{\cal M}}
\newcommand\cN{{\cal N}}
\newcommand\cG{{\cal G}}
\newcommand\cH{{\cal H}}
\newcommand\cI{{\cal I}}
\newcommand\cJ{{\cal J}}
\newcommand\cl{{\iota}}
\newcommand\cP{{\cal P}}
\newcommand\cQ{{\cal Q}}
\newcommand\cg{{\it g}}
\newcommand\cR{{\cal R}}
\newcommand\cB{{\cal B}}
\newcommand\cO{{\cal O}}
\newcommand\tcO{{\tilde {{\cal O}}}}
\newcommand\bz{\bar{z}}
\newcommand\bc{\bar{c}}
\newcommand\bw{\bar{w}}
\newcommand\bX{\bar{X}}
\newcommand\bK{\bar{K}}
\newcommand\bA{\bar{A}}
\newcommand\bZ{\bar{Z}}
\newcommand\bxi{\bar{\xi}}
\newcommand\bphi{\bar{\phi}}
\newcommand\bpsi{\bar{\psi}}
\newcommand\bprt{\bar{\prt}}
\newcommand\bet{\bar{\eta}}
\newcommand\btau{\bar{\tau}}
\newcommand\hF{\hat{F}}
\newcommand\hA{\hat{A}}
\newcommand\hT{\hat{T}}
\newcommand\htau{\hat{\tau}}
\newcommand\hD{\hat{D}}
\newcommand\hf{\hat{f}}
\newcommand\hg{\hat{g}}
\newcommand\hp{\hat{\phi}}
\newcommand\hi{\hat{i}}
\newcommand\ha{\hat{a}}
\newcommand\hb{\hat{b}}
\newcommand\hQ{\hat{Q}}
\newcommand\hP{\hat{\Phi}}
\newcommand\hS{\hat{S}}
\newcommand\hX{\hat{X}}
\newcommand\tL{\tilde{\cal L}}
\newcommand\hL{\hat{\cal L}}
\newcommand\tG{{\widetilde G}}
\newcommand\tg{{\widetilde g}}
\newcommand\tphi{{\widetilde \phi}}
\newcommand\tPhi{{\widetilde \Phi}}
\newcommand\te{{\tilde e}}
\newcommand\tk{{\tilde k}}
\newcommand\tf{{\tilde f}}
\newcommand\ta{{\tilde a}}
\newcommand\tb{{\tilde b}}
\newcommand\tR{{\tilde R}}
\newcommand\teta{{\tilde \eta}}
\newcommand\tF{{\widetilde F}}
\newcommand\tK{{\widetilde K}}
\newcommand\tE{{\widetilde E}}
\newcommand\tpsi{{\tilde \psi}}
\newcommand\tX{{\widetilde X}}
\newcommand\tD{{\widetilde D}}
\newcommand\tO{{\widetilde O}}
\newcommand\tS{{\tilde S}}
\newcommand\tB{{\widetilde B}}
\newcommand\tA{{\widetilde A}}
\newcommand\tT{{\widetilde T}}
\newcommand\tC{{\widetilde C}}
\newcommand\tV{{\widetilde V}}
\newcommand\thF{{\widetilde {\hat {F}}}}
\newcommand\Tr{{\rm Tr}}
\newcommand\tr{{\rm tr}}
\newcommand\STr{{\rm STr}}
\newcommand\hR{\hat{R}}
\newcommand\M[2]{M^{#1}{}_{#2}}

\newcommand\bS{\textbf{ S}}
\newcommand\bI{\textbf{ I}}
\newcommand\bJ{\textbf{ J}}

\begin{titlepage}
\begin{center}

\vskip 2 cm
{\LARGE \bf    $O(D,D)$-constraint   on $D$-dimensional  \\  \vskip 0.25 cm  effective  actions 
 }\\
\vskip 1.25 cm
   Mohammad R. Garousi\footnote{garousi@um.ac.ir}

\vskip 1 cm
{{\it Department of Physics, Faculty of Science, Ferdowsi University of Mashhad\\}{\it P.O. Box 1436, Mashhad, Iran}\\}
\vskip .1 cm
 \end{center}

\begin{abstract}
   Double Field Theory is a manifestly T-duality invariant formulation of string theory in which   the effective theory at any order of $\alpha'$  is  invariant under global $O(D,D)$    transformations and ought to be invariant under gauge transformations  
	which receive $\alpha'$-corrections.    
	On the other hand, the effective theory in the usual $D$-dimensional formulation of string theory is   manifestly gauge invariant and ought to be invariant under T-duality transformations which receive $\alpha'$-corrections. We speculate that the combination of these two constraints may fix  both  the $2D$-dimensional and the $D$-dimensional  effective actions without knowledge of  the $\alpha'$-corrections of the gauge   and   the  T-duality transformations.

In this paper,  using generalized fluxes,  we construct     arbitrary    $O(D,D)$-invariant   actions  at orders $\alpha'^0$ and   $\alpha'$, and then  dimensionally reduce them to the $D$-dimensional spacetime. On the other hand, at these orders,  we construct     arbitrary  covariant $D$-dimensional actions.       
 Constraining the two $D$-dimensional actions to be equal up to  non-covariant field redefinitions, we find that both    actions are fixed up to  overall factors and up to   field redefinitions.  
\end{abstract}
\end{titlepage}

\section{Introduction}

One of the most exciting discoveries in  string theory is   T-duality \cite{Giveon:1994fu,Alvarez:1994dn}. This   duality   may be used to construct the effective field theory at low energy. One approach for constructing this effective action  is the   Double Field Theory  (DFT)  approach  \cite{Siegel:1993xq,Siegel:1993th,Siegel:1993bj,Hull:2009mi,Aldazabal:2013sca}. The DFT is a constraint field theory which  doubles the spacetime coordinates, \ie adds to the usual $D$-dimensional spacetime coordinates which correspond to the momentum excitations,   another $D$-dimensional coordinates which correspond to the winding excitations. However, a solution to the constraint  in its strong form \cite{Aldazabal:2013sca}  is that the $2D$-dimensional dynamical fields to be   independent of the winding coordinates. The T-duality is manifested in this approach as the effective action is  $O(D,D)$-invariant by constructions. The effective action is also constrained to satisfy some gauge transformations. The appropriate gauge transformations at the leading order of $\alpha'$  are the generalized diffeomorphisms  and double-Lorentz transformations \cite{Hohm:2010pp,Aldazabal:2013sca}, however,  one of them  receive $\alpha'$-corrections at   the higher orders of $\alpha'$ \cite{Hohm:2014xsa,Marques:2015vua}.  The form of these corrections at order $\alpha'$ have been found in \cite{Hohm:2014xsa,Marques:2015vua}, however, it is hard to find them at the higher orders.   Using the $2D$-dimensional field redefinitions freedom, the effective action may appear in different schemes. The DFT effective action at order $\alpha'$ in one particular scheme has been constructed in \cite{Marques:2015vua,Baron:2017dvb}. 

Another  T-duality based approach for constructing the   effective action at higher orders of $\alpha'$, is to use the constraint that   the dimensional reduction of the effective action   on a circle  must be invariant under the T-duality transformations \cite{Garousi:2017fbe}. In this   approach, one begins with the most general gauge invariant  action in the  $D$-dimensional spacetime. The dimensional reduction of this action on the circle must be invariant under the   T-duality transformations. The gauge transformations in this approach are the standard coordinate transformations,  the  B-field gauge transformations and the non-standard Lorentz transformation of the B-field which is required for anomaly cancellations.   The T-duality transformations at  the leading order of $\alpha'$ are the Buscher rules \cite{Buscher:1987sk,Buscher:1987qj}, however, they receive $\alpha'$-corrections at   the higher orders of $\alpha'$ \cite{Bergshoeff:1995cg,Kaloper:1997ux}. The form of these corrections at order $\alpha'$ have been found in \cite{Tseytlin:1991wr,Bergshoeff:1995cg,Kaloper:1997ux}\footnote{It has been observed in  \cite{Haagensen:1997er} that  the renormalization group  flows  is covariant under the  T-duality  transformations at order $\alpha'$. }, however, it is hard to find them at higher orders of $\alpha'$.     Using the T-duality approach, the standard gravity and dilaton  couplings in the effective actions at orders $\alpha',\alpha'^2,\alpha'^3$ have been reproduced  in \cite{Razaghian:2017okr,Razaghian:2018svg}. 

Since the higher derivative corrections to the gauge transformations in the DFT approach and the $\alpha'$-corrections to the Buscher rules in the T-duality approach are hard to find in general, it is desirable to find   constraints   which do not receive $\alpha'$-corrections.   Merging the above two approaches, one may finds such constraints as follows: 
Using the strong constraint, one can reduce the $2D$-dimensional  effective action in the DFT approach   to the $D$-dimensional effective action.  This action should then be the same as the $D$-dimensional effective action in the T-duality approach. The field variables in the two approaches, however, are not the same. Non-covariant field redefinitions are required to relate the two field variables \cite{Meissner:1996sa}. The T-duality transformations in the DFT approach are the standard Buscher rules \cite{Aldazabal:2013sca} whereas   the gauge transformations   are not the standard gauge transformations. On the other hand, the gauge transformations in the T-duality approach are the standard gauge transformations whereas the T-duality transformations are not the standard Buscher rules. As a result, the two $D$-dimensional actions must be the same up to  non-covariant field redefinitions. Therefore, the effective actions should satisfy the following constraints: 

1-The $2D$-dimensional action is constrained to be invariant under the $O(D,D)$ transformations and under the  generalized diffeomorphisms which do not receive $\alpha'$-corrections. However, it is not constrained to be invariant under the double-Lorentz transformations which receive $\alpha'$-corrections.

2-After reducing it to the $D$-dimensional spacetime and using non-covariant field redefinitions, the action is constrained to be the same as a $D$-dimensional action which is invariant under the standard coordinate transformations,  the  B-field gauge transformations and the non-standard Lorentz transformation of the B-field, however, it is not   invariant under the T-duality transformations which receive $\alpha'$-corrections.
We speculate that the above two constraints can fix both the $2D$-dimensional and $D$-dimensional effective actions. We confirm this idea    in this paper by explicit calculations at orders $\alpha'^0$ and $\alpha'$.

The outline of the paper is as follows: In section 2, we pefrom the calculations at order $\alpha'^0$.   In particular, in subsection 2.1, we use generalized metric and dilaton as dynamical fields which are invariant under the double-Lorentz transformations, to construct the most general $O(D,D)$-invariant action at order $\alpha'^0$. Using the strong constraint, we then reduce it to the $D$-dimensional action. Then, using the $D$-dimensional metric, B-field and dilaton, we construct the most general  covariant action at order $\alpha'^0$. Constraining the two actions to be identical, we fix both   effective actions. Up to an overall factor, they are exactly the known effective actions in the literature. In section 2.2, we use the generalized frame and dilaton as the dynamical fields. Using the generalized fluxes, which are invariant under the generalized diffeomorphisms, we construct the most general $O(D,D)$-invariant action at order $\alpha'^0$, and then reduce it to the $D$-dimensional action. Comparing it with the  covariant $D$-dimensional action, we   fix both  the  effective actions. The $2D$-dimensional effective action is the same as the action in the literature.  

In section 3, we extend the calculations to the order $\alpha'$. In particular, using the generalized fluxes, we first construct the most general $O(D,D)$-invariant action at order $\alpha'$ without fixing its field redefinitions freedom, and then reduce it to the $D$-dimensional action. To convert the non-covariant field variables in the resulting action to the covariant variables, we   use  the most general    non-covariant field redefinitions. We then compare it with  the most general covariant action at order $\alpha'$ up to covariant field redefinitions. The constraint that  the two $D$-dimensional actions must be identical, fixes both actions. Up to an overall factor, the $D$-dimensional action is exactly the same as the action in the literature. Since the field redefinitions freedom is not fixed in the $2D$-dimensional action, we have found the $2D$-dimensional action with some arbitrary parameters. In one particular scheme in which dilaton appears as an overall factor, we write the effective action.

\section{Effective action at order $\alpha'^0$}

Using the strong constraint in the DFT formalism,  the effective action  of string theory at order $\alpha'^0$ can be written as $O(D,D)$-invariant and invariant under $2D$-dimensional gauge transformations which are generalized diffeomorphisms and local double-Lorentz transformations. If one uses the  generalized metric and dilaton as dynamical fields which are invariant under the double-Lorentz transformations, then the gauge transformations are the  generalized diffeomorphisms \cite{Hohm:2010pp}. On the other hand, if one uses the generalized frame and dilaton as the dynamical fields, then the action can be written in terms of  generalized fluxes which are invariant under the generalized diffeomorphisms \cite{Aldazabal:2013sca}. Hence, the nontrivial gauge transformations in this case is the double-Lorentz transformations. Using these gauge transformations, the $2D$-dimensional effective actions have been found in \cite{Hohm:2010pp,Geissbuhler:2011mx}. 

 In this section we are going to find these actions by comparing the most general $O(D,D)$-invariant action with the most general  $D$-dimensional covariant action.

\subsection{Generalized metric formulation}

We begin with the case that the generalized metric  ${\cal H}_{\mu\nu}$  and dilaton $d$ are the dynamical fields. They are invariant under the double-Lorentz transformations as they carry no index in this space, however,   the generalized metric ${\cal H}_{\mu\nu}$ is a matrix that transforms under the   $O(D,D)$ transformations as\footnote{Our index conversion is that the Greek letters  $(\mu,\nu,\cdots)$ are  the indices of the curved $2D$-dimensional space, the Latin letters $(a,d,c,\cdots)$ are  the indices of the curved $D$-dimensional spacetime,  the letters $(A,B,C, \cdots)$ are the indices of flat   $2D$-dimensional tangent space,   and the letters $(i,j,k,\cdots)$ are the flat $D$-dimensional tangent space.}
\beqa
 \cH\rightarrow O\cH O^T
\eeqa 
 The $D$-dimensional coordinate $x^a$ conjugated   to the momentum excitations and the $D$-dimensional coordinate $\tilde{x}_a$ conjugated to the winding excitations, transforms as vector, \ie
\beqa
x^{\mu}\equiv\pmatrix {\tilde{x}_a\cr x^a}\rightarrow O\pmatrix {\tilde{x}_a\cr x^a}\,\,\,;\,\,\,\prt_{\mu}\equiv\pmatrix {\tilde{\prt}^a\cr \prt_a}\rightarrow O\pmatrix {\tilde{\prt}^a\cr \prt_a}
\eeqa
And the $2D$-dimensional dilaton $d$ is invariant under the   $O(D,D)$ transformations \cite{Hohm:2010pp}. Using these $O(D,D)$ tensors,  one can write the most general  $O(D,D)$-invariant action at two-derivative level as\footnote{We use the mathematica package 'xAct' \cite{Nutma:2013zea} for performing the calculations in this paper.} 
\beqa
\bS_0&=&-\frac{2 }{\kappa^2 \tV}\int dx d\tilde{x} e^{-2d}\Big( c_9  {\cal H}^{\alpha \beta} \partial_{\alpha}d \partial_{\beta}
d + c_3  {\cal H}^{\alpha \beta} 
\partial_{\beta}\partial_{\alpha}d + c_1  {\cal H}^{\alpha 
\gamma} {\cal H}^{\beta \delta} \partial_{\gamma}\partial_{\alpha}{\cal H}_{\delta 
\beta}\nn\\&&\qquad\quad + c_2  {\cal H}^{\alpha \gamma} {\cal H}^{\beta \delta} 
\partial_{\gamma}\partial_{\delta}{\cal H}_{\alpha \beta}  + c_5  
{\cal H}^{\alpha \delta} {\cal H}^{\beta \epsilon} {\cal H}^{\gamma \varepsilon} 
\partial_{\alpha}{\cal H}_{\beta \gamma} \partial_{\delta}{\cal H}_{\epsilon 
\varepsilon} + c_6  {\cal H}^{\alpha \delta} {\cal H}^{\beta \epsilon} {\cal H}^{
\gamma \varepsilon} \partial_{\beta}{\cal H}_{\alpha \gamma} 
\partial_{\delta}{\cal H}_{\epsilon \varepsilon} \nn\\&&\qquad\quad + c_{11}  
{\cal H}^{\alpha \delta} {\cal H}^{\beta \gamma} \partial_{\gamma}{\cal H}_{\alpha \beta} 
\partial_{\delta}d + c_8  {\cal H}^{\alpha \delta} {\cal H}^{\beta 
\epsilon} {\cal H}^{\gamma \varepsilon} \partial_{\delta}{\cal H}_{\alpha \epsilon} 
\partial_{\varepsilon}{\cal H}_{\beta \gamma}+ c_{10} 
{\cal H}^{\alpha \delta} {\cal H}^{\beta \gamma} \partial_{\alpha}{\cal H}_{\gamma 
\beta} \partial_{\delta}d\nn\\&&\qquad\quad 
+c_7  {\cal H}^{\alpha \delta} {\cal H}^{\beta \epsilon} {\cal H}^{\gamma 
\varepsilon} \partial_{\beta}{\cal H}_{\varepsilon \gamma} 
\partial_{\delta}{\cal H}_{\alpha \epsilon}  + c_4  {\cal H}^{\alpha 
\epsilon} {\cal H}^{\beta \varepsilon} {\cal H}^{\gamma \delta} 
\partial_{\alpha}{\cal H}_{\delta \gamma} 
\partial_{\epsilon}{\cal H}_{\varepsilon \beta}
\Big)\labell{ODD}
\eeqa
where ${\cal H}^{\mu\nu}$ is inverse of the generalized metric and $\tV=\int d\tilde{x}$ . Since there is no double-Lorentz index in the couplings \reef{ODD}, this action is invariant under the local double-Lorentz transformations as well.

There is also a  $O(D,D)$-invariant metric 
\beqa
{\eta}_{\mu\nu}=\pmatrix{ 0&1\cr1 &0 }&;&{\eta}^{\mu\nu}=\pmatrix{0&1\cr 1&0}\labell{eta}
\eeqa
which raises and lowers the $2D$-indicies, \ie
\beqa
\cH^{\mu}{}_{\nu}=\eta^{\mu\alpha}\cH_{\alpha\nu}&;&{\cal H}^{\mu\nu}=\eta^{\mu\alpha}\eta^{\nu\beta}\cH_{\alpha\beta}\,\,\,;\,\,\prt^\alpha=\eta^{\alpha\beta}\prt_\beta
\eeqa 
The symmetry of the effective action under $B_{ab}\rightarrow -B_{ab}$, requires the couplings   to have     even number of constant metric $\eta$ \cite{Hohm:2010pp}.  The couplings involving odd number of $\eta$ are antisymmetric under  $B_{ab}\rightarrow -B_{ab}$. At two derivative level, one can convince oneself that any term  which is independent of the winding coordinates and contains  two constant metrics, is identical to the couplings in \reef{ODD}. For example, using the constant metric to  raise and lower the indices, the term $\prt_{\alpha} \cH^\alpha{}_{\beta}\prt_\gamma\cH^{\gamma}{}_{\mu} \cH^{\beta\mu}$ can be written as  $\prt_{\alpha} \cH^{\alpha \beta}\prt_\gamma\cH^{\gamma \mu} \cH_{\beta\mu}$. The latter coupling can be written as the couplings in \reef{ODD} using the identity 
\beqa
 {\cal H}^{\alpha\beta}{\cal H}_{\beta\mu}&=&\delta ^{\alpha}_{\mu}\label{iden}
\eeqa
At higher derivative level, however, there are  couplings involving the metric $\eta$ which can not be written in terms of only generalized metric, \eg $\prt_{\alpha}\cH^{\alpha}{}_{\beta}\prt_\gamma\cH^\gamma{}_\lambda\prt_\mu\cH^{\mu\lambda}\prt_\nu\cH^{\nu\beta}$. 

The couplings in \reef{ODD} with coefficients $c_4,c_7,c_{10}$ become  zero using the above identity. Using this identity and total derivative terms,  one can relate the coefficients of some of the above terms. One may either use these relations to write \reef{ODD} in terms of independent couplings and then compare them with  the $D$-dimensional gauge invariant action, or one may   fix them after comparing the  non-independent couplings  with the $D$-dimensional   gauge invariant action. In the latter case that we are going to do in this paper,  they appear as free parameters which can be chosen arbitrarily.  

The reduction of the generalized metric and its inverse in terms of the $D$-dimensional metric and the B-field are \cite{Hohm:2010pp}
\beqa
{\cal H}_{\mu\nu}=\pmatrix{ G^ {ab}&-G^{ac}B_{cb}\cr B_{ac}G^ {cb} &G_{ab}-B_{ac}G^ {cd}B_{db}  }&;&{\cal H}^{\mu\nu}=\pmatrix{G_{ab}-B_{ac}G^{cd}B_{db}&B_{ac}G^{cb}\cr -G^{ac}B_{cb}&G^{ab}}\labell{genmet}
\eeqa
The reduction of the $2D$-dimensional dilaton to the $D$-dimensional dilaton and metric is $e^{-2d}= e^{-2\Phi}\sqrt{-G}$. Using the strong constraint that fields   do not depend on the coordinate $\tilde{x}$, one can reduce \reef{ODD}  to the  following  $D$-dimensional action: 
\beqa
 \bS_0&=&-\frac{2}{\kappa^2}\int dx  e^{-2\Phi}\sqrt{-G}\Big( c_9  G^{ab} \partial_{a}\Phi \partial_{b}\Phi + 
c_3  G^{ab} \partial_{b}\partial_{a}\Phi -  
 \frac{1}{4} c_3  G^{ac} G^{bd} 
\partial_{c}\partial_{a}G_{db} \nn\\&&\qquad\qquad\qquad + 2 (c_1  + 
c_5 ) G^{ad} G^{be} G^{cf} \partial_{a}B_{bc} 
\partial_{d}B_{ef} + (c_2  + c_6 ) G^{ad} 
G^{be} G^{cf} \partial_{b}B_{ac} \partial_{d}B_{ef}\nn\\&&\qquad\qquad\qquad -  
 \frac{1}{4} c_{11}  G^{ad} G^{be} G^{cf} 
\partial_{b}G_{fc} \partial_{d}G_{ae} +  \frac{1}{4} (8 
c_1  + c_3  + 8 c_5 ) G^{ad} G^{be} 
G^{cf} \partial_{a}G_{bc} \partial_{d}G_{ef}\nn\\&&\qquad\qquad\qquad + c_6  
G^{ad} G^{be} G^{cf} \partial_{b}G_{ac} \partial_{d}G_{ef} -  
 \frac{1}{2} c_9  G^{ad} G^{bc} \partial_{a}G_{cb} 
\partial_{d}\Phi + c_{11}  G^{ad} G^{bc} 
\partial_{c}G_{ab} \partial_{d}\Phi\nn\\&&\qquad\qquad\qquad +  \frac{1}{16} 
c_9  G^{ae} G^{bf} G^{cd} \partial_{a}G_{dc} 
\partial_{e}G_{fb} + (- c_2  -  c_8 ) G^{ad} 
G^{be} G^{cf} \partial_{d}B_{ae} \partial_{f}B_{bc}\nn\\&&\qquad\qquad\qquad + 
c_8  G^{ad} G^{be} G^{cf} \partial_{d}G_{ae} 
\partial_{f}G_{bc}+ c_2  G^{ac} G^{bd} 
\partial_{c}\partial_{d}G_{ab} \Big)\labell{S01}
 \eeqa
Note that the coefficients $c_4,c_7,c_{10}$ do  not appear in above $D$-dimensional action, so they appear in the final action  as arbitrary parameters which can be set to zero\footnote{ One can check that the above couplings are invariant under the Buscher rules, \ie if one compactifies the theory on a circle and assumes fields are independent of that directions, then the above couplings would be  invariant under the Buscher rules.}.   It is interesting to note that even though the generalized metric contains no derivative on the B-field, the above $D$-dimensional action contains only terms which have derivative on the B-field. It turns out that the dimensional reduction of any $O(D,D)$-invariant coupling produces no B-field without derivative on it.
The above  $O(D,D)$ and double-Lorentz invariant  action  is not invariant under the generalized diffeomorphism  for arbitrary parameters. We do not impose this constraint for finding the unknown coefficients in this paper. 

 We now construct   the most general $D$-dimensional  action at two-derivative level which is invariant under the coordinate transformations and under the B-field gauge transformations, \ie
\beqa
\bS^c_0&=& -\frac{2}{\kappa^2}\int dx e^{-2\Phi}\sqrt{-G}\,  \left(a_1 R + a_2\nabla_{a}\Phi \nabla^{a}\Phi-\frac{a_3}{12}H^2\right)\,.\labell{S0b}
\eeqa
where $H_{abc}=\prt_a B_{bc}+\prt_c B_{ab}+\prt_b B_{ca}$ and $a_1,a_2,a_3$ are three constants. Since the action \reef{S01} is in terms of metric  and B-field, we rewrite the above covariant action in terms of metric and B-field, \ie
\beqa
\bS^c_0&=& -\frac{2}{\kappa^2}\int dx e^{-2\Phi}\sqrt{-G}\Big(a_2 G^{ab} \partial_{a}\Phi \partial_{b}\Phi -  a_1 G^{ac} 
G^{bd} \partial_{c}\partial_{a}G_{db} + a_1 G^{ac} G^{bd} 
\partial_{c}\partial_{d}G_{ab}\nn\\&&\qquad\qquad\qquad -   \frac{1}{4} a_3 G^{ad} G^{be} 
G^{cf} \partial_{a}B_{bc} \partial_{d}B_{ef} +  \frac{1}{2} a_3 
G^{ad} G^{be} G^{cf} \partial_{b}B_{ac} \partial_{d}B_{ef} \nn\\&&\qquad\qquad\qquad+ 
a_1 G^{ad} G^{be} G^{cf} \partial_{b}G_{fc} \partial_{d}G_{ae} 
+  \frac{3}{4} a_1 G^{ad} G^{be} G^{cf} \partial_{a}G_{bc} 
\partial_{d}G_{ef} \nn\\&&\qquad\qquad\qquad-   \frac{1}{2} a_1 G^{ad} G^{be} G^{cf} 
\partial_{b}G_{ac} \partial_{d}G_{ef} -   \frac{1}{4} a_1 G^{ae} 
G^{bf} G^{cd} \partial_{a}G_{dc} \partial_{e}G_{fb} \nn\\&&\qquad\qquad\qquad-  a_1 
G^{ad} G^{be} G^{cf} \partial_{d}G_{ae} \partial_{f}G_{bc}\Big)\labell{S01c}
\eeqa
The two $D$-dimensional Lagrangians \reef{S01} and \reef{S01c} are not equal for any non-zero parameters. However, to compare the two actions, one should  take into account non-covariant total derivative terms as well. There are three total derivative terms, \ie  
\beqa
\bJ&=& -\frac{2}{\kappa^2}\int dx \prt_a\Big(e^{-2\Phi}\sqrt{-G} [f_3  G^{ab} \partial_{b}\Phi + f_2  G^{cb} 
G^{da} \partial_{c}G_{db} + f_1  G^{ca} G^{db} 
\partial_{c}G_{db}]\Big)
\eeqa
where $f_1,f_2,f_3$ are three arbitrary parameters. Now adding these total derivative terms to \reef{S01c}, the two actions can be equated, \ie $\bS_0=\bS_0^c+\bJ$, for the following constraints on the parameters: 
\beqa
&&a_2 = 4 a_1, a_3 = a_1,\nn\\&& c_3 = 2 a_1 - c_{11} - 2 c_2, c_5 = -a_1/8 - c_1, 
 c_6 = a_1/2 - c_2, c_8 = -c_2, c_9 = 2 c_{11} + 4 c_2,\nn\\&& 
 f_1 = a_1/2 + c_{11}/4 + c_2/2, f_2 = -a_1 + c_2, f_3 = 2 a_1 - c_{11} - 2 c_2
\eeqa
The equations in the last line gives the coefficients of the total derivative terms that are needed to equate the two actions.

The equations in  the first line constrain the $D$-dimensional action \reef{S0b} to be 
\beqa
\bS^c_0&=& -\frac{2a_1}{\kappa^2}\int dx e^{-2\Phi}\sqrt{-G}\,  \left(  R + 4\nabla_{a}\Phi \nabla^{a}\Phi-\frac{1}{12}H^2\right)\,.\labell{S0bf}
\eeqa
which is the standard effective action at order $\alpha'^0$, up to an overall factor. The overall factor must be $a_1=1$ to be the effective action of string theory.
The equations in the second line constrain the $O(D,D)$-invariant  action \reef{ODD} to be 
\beqa
\bS_0& = &-\frac{2a_1}{\kappa^2\tV} \int dx d\tilde{x} e^{-2d}\Big(   2{\cal H}^{\alpha \beta} 
\partial_{\beta}\partial_{\alpha}d    +\frac{1}{8}
{\cal H}^{\alpha \delta}   
\partial_{\alpha}{\cal H}_{\beta \gamma} \partial_{\delta}{\cal H}^{\beta\gamma} -\frac{1}{2} {\cal H}^{\alpha \delta}   \partial_{\beta}{\cal H}_{\alpha \gamma} 
\partial_{\delta}{\cal H}^{\beta\gamma}  
\Big)\labell{ODD1}
\eeqa
where we have also used the identity $\prt_{\alpha}{\cal H}^{\mu\nu}=-{\cal H}^{\mu\beta}{\cal H}^{\nu\gamma}\prt_{\alpha}{\cal H}_{\beta\gamma}$. The terms with coefficients $c_2,c_{11}$ are total derivative terms, and terms with coefficient $c_1$, \ie
\beqa
 c_1  {\cal H}^{\gamma \alpha} {\cal H}^{\delta \beta} 
\partial_{\gamma}\partial_{\alpha}{\cal H}_{\delta \beta} -  
c_1  {\cal H}^{\delta \alpha} {\cal H}^{\epsilon \beta} 
{\cal H}^{\varepsilon \gamma} \partial_{\alpha}{\cal H}_{\beta \gamma} 
\partial_{\delta}{\cal H}_{\epsilon \varepsilon}
\eeqa
become zero using the identity  \reef{iden}, so we have discarded them.  The $O(D,D)$-invariant action \reef{ODD1} is the one has been found in \cite{Hohm:2010pp}. Therefore, the requirement that the   $O(D,D)$-invariant action and the  covariant   $D$-dimensional action to be identical, fixes both actions, up to an overall factor. 

The action \reef{ODD1} has been found in \cite{Hohm:2010pp} by requiring the  $O(D,D)$-invariant couplings \reef{ODD} to be invariant under the generalized diffeomorphisms which are
\beqa
\delta (e^{-2d})&=&\prt_\mu(\z^\mu e^{-2d})\nn\\
\delta {\cal H}_{\mu\nu}&=& \z^\rho\prt_\rho{\cal H}_{\mu\nu}+2(\prt_{(\mu}\z^\rho-\prt^\rho\z_{(\mu}){\cal H}_{\nu)\rho}\labell{gcov}
\eeqa
Unlike the terms in \reef{S0b} which are invariant under the $D$-dimensional diffeomorphisms, none of the terms in \reef{ODD} is invariant under the above $2D$-dimensional generalized diffeomorphisms. Only the combination of terms in \reef{ODD1} is invariant under these transformations \cite{Hohm:2010pp}. This combination may be defined as the definition of $2D$-dimensional scalar curvature \cite{Hohm:2010pp}. 
It is hard to extend the couplings \reef{ODD1} to the higher order of $\alpha'$ because the conventional  $2D$-dimensional Riemann curvature does not transform covariantly  under the generalized diffeomorphisms \cite{Siegel:1993th,Hohm:2010xe,Hohm:2011si,Jeon:2011cn,Coimbra:2011nw}.

\subsection{Generalized frame formulation}

A convenient frame work for constructing the higher derivative couplings in DFT is the generalized frame construction of the DFT \cite{Siegel:1993xq,Aldazabal:2013sca}. The generalized frame $E_{\mu}{}^A$ is defined to relate  the generalized metric $\cH_{\mu\nu}$ to the flat generalized metric $\cH_{AB}$ and the $O(D,D)$-metric $\eta_{\mu\nu}$ to flat metric $\eta_{AB}$, \ie
\beqa
 \cH_{\mu\nu}&=&E_{\mu}{}^A\cH_{AB}E_{\nu}{}^B\nn\\
\eta_{\mu\nu}&=&E_{\mu}{}^A\eta_{AB}E_{\nu}{}^B
\eeqa
In terms of the generalized dilaton $d$,   the generalized frame $E_{\mu}{}^A$ and its transverse $E^{\mu}{}_A$, one can construct flat space tensors which transform as scalar under the generalized diffeomorphisms. They are 
\beqa
\cF_A&=&2\prt_A d-E^{\mu B}\prt_B E_{\mu A}\nn\\
\cF_{ABC}&=&3\prt_{[A}E^{\mu}{}_B E^{\nu}{}_{C]}\eta_{\mu\nu}\labell{FF}
\eeqa
where the flat space derivative is $\prt_A=E^{\mu}{}_A\prt_{\mu}$. These tensors transform as scalar under the generalized diffeomorphisms \cite{Aldazabal:2013sca}, \ie
\beqa
\delta \cF_{A}&=& \z^\mu\prt_\mu \cF_A\nn\\
\delta \cF_{ABC}&=&\z^\mu\prt_{\mu} \cF_{ABC}
\eeqa
The flat space derivatives of these tensors transform as scalar under the generalized diffeomophisms as well \cite{Aldazabal:2013sca}. However, these tensors do not transform covariantly under local double-Lorentz transformations. It has been shown in \cite{Marques:2015vua,Baron:2017dvb} that these transformations receive $\alpha'$ corrections as well\footnote{If one uses the generalized metric and dilaton as the dynamical fields which are invariant under the double-Lorentz transformations, then the $\alpha'$-corrections would appear in the generalized diffoemorphisms \cite{Hohm:2014xsa}.}. 

One may construct $O(D,D)$-invariant and the generalized diffeomophism invariant effective actions by considering all contractions of these tensors with constant metric $\eta^{AB}$ and $\cH^{AB}$, \ie at two-derivative level they are
\beqa
\bS_0^f&=&-\frac{2}{\kappa^2\tV}\int dx d\tilde{x} e^{-2d}\Big(c_3 \cF_{A} \cF^{A} + c_4 
\prt^A\cF_{A} + c_1 \cF^{A} 
\cF^{B} \cH_{A B} + c_2 \prt^A\cF^{  
B} \cH_{A B} + c_7 \cF_{A B C} \cF^{A B C}\nn\\&& + 
c_6 \cH^{A B} \cF_{A}{}^{CD} \cF_{B CD} + 
c_5 \cH^{A B} \cH^{CD} 
\cF_{A C}{}^{E} \cF_{B 
DE} + c_8 \cH^{A B} 
\cH^{CD} \cH^{E F} \cF_{A C E} \cF_{BD 
F}\Big)\labell{S0F}
\eeqa
The flat indices are raised by the flat metric $\eta^{AB}$, \ie $\cF^A=\eta^{AB}\cF_{B}$. This action is invariant under the generalized diffeomophisms for arbitrary parameters $c_1,\cdots, c_8$, however, it is not invariant under the local double-Lorentz transformations. Imposing the invariance   under the double-Lorentz transformations,  one can fix these parameters \cite{Geissbuhler:2011mx,Baron:2017dvb}. However, we are not going to fix the parameters in this way.

To fix the  parameters $c_1,\cdots, c_8$, instead, we constrain the  reduction of this action to be identical with the $D$-dimensional covariant action \reef{S0b}. 
The reduction of the metric $\eta^{AB}$, $\cH^{AB}$ and the  generalized frame $E_\mu{}^A$ are \cite{Baron:2017dvb}: 
\beqa
\eta^{AB}=\pmatrix{0&\delta_i^j\cr \delta^i_j &0 }&;&{\cal H}^{AB}=\pmatrix{\eta_{ij}&0\cr0&\eta^{ij}}\,\,;\,\,E_{\mu}{}^A=\pmatrix{e^a{}_i&0\cr-e^b{}_iB_{ba}&e_a{}^i}\labell{genmet1}
\eeqa
where $e_a{}^i$ is the $D$-dimensional frame, \ie $e_a{}^i e_b^j\eta_{ij}=G_{ab}$.  Using the   constraint that fields in  the  $2D$-dimensional action \reef{S0F} do not depend on the coordinate $\tilde{x}$, one can reduce it  to the  following  $D$-dimensional action:
\beqa
\bS^f_0&=&-\frac{2}{\kappa^2} \int dx e^{-2\Phi}\sqrt{-G}\Big(c_2  G^{ad} G^{bc}  \eta_{j i} e_{d}{}^{ i} 
\partial_{a}\partial_{c}e_{b}{}^{j} + 4 c_1  G^{ab} 
\partial_{a}\Phi \partial_{b}\Phi + 2 c_2  G^{ab} 
\partial_{b}\partial_{a}\Phi\nn\\&& + 6 c_8  G^{ac} G^{bd} 
 \eta_{j i} \partial_{a}e_{b}{}^{j} \partial_{c}e_{d}{}^{ i} - 
4 c_1  G^{ac} G^{bd}  \eta_{j i} e_{b}{}^{j} 
\partial_{a}\Phi \partial_{c}e_{d}{}^{ i}- 6 c_8  
G^{ac} G^{bd}  \eta_{j i} \partial_{b}e_{a}{}^{j} 
\partial_{c}e_{d}{}^{ i} \nn\\&& + 4 c_1  G^{ac} G^{bd} \eta_{j i} e_{a}{}^{j} \partial_{b}\Phi \partial_{c}e_{d}{}^{ i} -  c_2  G^{ad} G^{bc} \eta_{j i} e_{d}{}^{ i} \partial_{c}\partial_{b}e_{a}{}^{j} + 3 c_8  G^{ad} G^{be} G^{cf} \partial_{a}B_{bc} \partial_{d}B_{ef} \nn\\&& - 6 c_8  G^{ad} G^{be} G^{cf} \partial_{b}B_{ac} \partial_{d}B_{ef} + 4 c_5  G^{ad} G^{be} G^{cf}  \eta_{j i} e_{f}{}^{j} \partial_{a}B_{bc} \partial_{d}e_{e}{}^{ i} \nn\\&& - 4 c_5  G^{ad} G^{be} G^{cf}  \eta_{j i} e_{f}{}^{j} \partial_{b}B_{ac} \partial_{d}e_{e}{}^{ i} + (- c_2  + 4 c_6 ) G^{ad} G^{be} G^{cf}  \eta_{kj}  \eta_{l i} e_{a}{}^{k} 
e_{c}{}^{l} \partial_{b}e_{f}{}^{j} \
\partial_{d}e_{e}{}^{ i}\nn\\&&  - 2 c_1  G^{ad} G^{be} 
G^{cf}  \eta_{k i}  \eta_{lj} e_{a}{}^{k} e_{c}{}^{l} 
\partial_{b}e_{f}{}^{j} \partial_{d}e_{e}{}^{ i} + 4 
c_5  G^{ad} G^{be} G^{cf}  \eta_{j i} e_{f}{}^{j} 
\partial_{c}B_{ab} \partial_{d}e_{e}{}^{ i} \nn\\&& - 2 c_2  
G^{ac} G^{bd}  \eta_{j i} e_{b}{}^{j} \partial_{c}e_{a}{}^{ i} 
\partial_{d}\Phi + (c_2  - 2 c_6 ) G^{ae} 
G^{bf} G^{cd}  \eta_{kj}  \eta_{l i} e_{b}{}^{k} e_{c}{}^{l} 
\partial_{a}e_{d}{}^{j} \partial_{e}e_{f}{}^{ i}\nn\\&&  + 
c_1  G^{ae} G^{bf} G^{cd}  \eta_{k i}  \eta_{lj} 
e_{b}{}^{k} e_{c}{}^{l} \partial_{a}e_{d}{}^{j} 
\partial_{e}e_{f}{}^{ i} -  c_2  G^{ad} G^{be} G^{cf} 
 \eta_{kj}  \eta_{l i} e_{b}{}^{k} e_{c}{}^{l} 
\partial_{d}e_{a}{}^{ i} \partial_{e}e_{f}{}^{j} \nn\\&& + 
c_2  G^{ad} G^{be} G^{cf}  \eta_{k i}  \eta_{lj} 
e_{b}{}^{k} e_{c}{}^{l} \partial_{d}e_{a}{}^{ i} 
\partial_{e}e_{f}{}^{j} - 2 c_6  G^{ad} G^{be} 
G^{cf}  \eta_{kj}  \eta_{l i} e_{a}{}^{k} e_{c}{}^{l} 
\partial_{d}e_{e}{}^{ i} \partial_{f}e_{b}{}^{j} \nn\\&& + 
c_1  G^{ad} G^{be} G^{cf}  \eta_{k i}  \eta_{lj} 
e_{a}{}^{k} e_{c}{}^{l} \partial_{d}e_{e}{}^{ i} 
\partial_{f}e_{b}{}^{j}\Big)
\eeqa
The terms with coefficients $c_3,c_4,c_7$, which have no $\cH^{AB}$,  vanish when reducing   the couplings \reef{S0F} to the $D$-dimensional spacetime. Hence, these terms are zero by the strong constraint. Note that even though the generalized frame contains no derivative on the B-field, the above $D$-dimensional action contains only terms which have derivative on the B-field.  The above $D$-dimensional action must be equal to \reef{S01c} plus some  $D$-dimensional total derivative terms. To compare the two actions, one has to rewrite the derivatives of metric in the action \reef{S01c} in terms of derivatives of the frame $e_a{}^i$. The comparison then fixes both the effective actions \reef{S0b} and \reef{S0F} up to an overall factor. The action \reef{S0b} is fixed as in \reef{S0bf} and the action \reef{S0F} is fixed as 
\beqa
\bS_0^f&=&-\frac{2a_1}{\kappa^2\tV}\int dx d\tilde{x} e^{-2d}\Big(  - \cF^{A} 
\cF^{B} \cH_{A B} + 2 \prt^A\cF^{  
B} \cH_{A B} \nn\\&& + 
\frac{1}{4} \cH^{A B} \cF_{A}{}^{CD} \cF_{B CD}   -\frac{1}{12} \cH^{A B} 
\cH^{CD} \cH^{E F} \cF_{A C E} \cF_{BD 
F}+\cdots\Big)\labell{S0Ff}
\eeqa
where dots represent the terms which vanish after using the strong constraint. The above action is    the action has been found in \cite{Geissbuhler:2011mx}.  In the next section we consider this approach to find both $D$-dimensional and $2D$-dimensional effective actions at order $\alpha'$.

\section{Effective action at order $\alpha'$}

The most general four-derivative action which is $O(D,D)$-invariant and is invariant under generalized diffeomorphisms can be constructed by all possible contractions of  the following tensors with constant metric $\eta^{AB}$ and $\cH^{AB}$:
\beqa
&&\cF_{A} \cF_{B} \cF_{C} \cF_{D} +   
\cF_{A} \cF_{B} \cF_{C} \cF_{D 
EF} +   \cF_{A} \cF_{B} \cF_{C D E} \cF_{F GH} +   \cF_{A} \cF_{B C D} \cF_{EF G} \cF_{H 
I J}\nn\\&& + \cF_{A B C} 
\cF_{D EF} \cF_{G 
H I} \cF_{J KL} +   
\cF_{C D E} \cF_{F 
GH} \partial_{A}\cF_{B} +   \cF_{
D EF} \partial_{A}\partial_{B}\cF_{C} +  \partial_{A}\partial_{B}\partial_{C}\cF_{D} \nn\\&&+   
\partial_{A}\partial_{B}\partial_{C}\cF_{D EF} +   \cF_{A} \cF_{D E F} \partial_{B}\cF_{C} +   \cF_{A} \partial_{B}\partial_{C}\cF_{D} +   \cF_{A} \partial_{
B}\partial_{C}\cF_{D E
F} +  \cF_{A} \cF_{B} \partial_{C}\cF_{D}\nn\\&& +  \partial_{A}\cF_{B} 
\partial_{C}\cF_{D} +   \cF_{A} \cF_{B} 
\partial_{C}\cF_{D EF} 
+   \partial_{A}\cF_{B} \partial_{C}\cF_{D EF} +   \cF_{A
B C} \partial_{D}\partial_{E}\cF_{F GH} +  \cF_{A} 
\cF_{B C D} \partial_{E}\cF_{F GH}\nn\\&& +   \partial_{A}\cF_{B C D} \partial_{E}\cF_{F GH} +  \cF_{A B C} \cF_{D EF} \partial_{G}\cF_{H I J}
\eeqa
It produces 275 terms, \ie
\beqa
\bS_1^f&=& -\frac{2 }{\kappa^2\tV}\int dx d\tilde{x} e^{-2d}\Big(c_1\cH^{AB}\cH^{CD}\cF_{A} \cF_{B} \cF_{C} \cF_{D}+c_2 \cF^{A} \cF_{A} \cF^{B} \cF_{B}+\cdots\Big)\labell{S1F}
\eeqa
where $c_1,\cdots, c_{275}$ are parameters. The 
  terms which have no tensor $\cH^{AB}$, \eg $c_2$-term, are   zero after using Bianchi identities \cite{Geissbuhler:2013uka} and the strong constraint. There are 34 such terms. There are also 91 terms with other structures that become zero under these  constraints\footnote{One may use the Bianchi identities found in \cite{Geissbuhler:2013uka} to demonstrate the vanishing of some of these 125 terms. However, the more convenient way to impose the Bianchi identities and the strong constraint is to reduce the couplings to the $D$ dimensions using the reduction \reef{genmet1} and then use the strong constraint. The above 125 terms vanish when we write the $D$-dimensional couplings in terms of independent expressions.}. The remaining 150 terms are in five classes. One class includes terms that cancel each others  after using the strong constraint, one class includes terms that are total derivatives, one class includes  terms that become  zero   using appropriate identities, one class includes terms that are reproduced by field redefinitions at order $\alpha'$ of the fields in \reef{S0Ff}, and the last class includes all other terms in which we are interested.  One may examine all terms in details to exclude all terms except  the terms in the last class, and then imposes the $D$-dimensional gauge symmetry. Alternatively, one may impose the $D$-dimensional gauge symmetry on all 150 terms. In this case,  the terms in the first four classes appear in the final action with free parameters which can be chosen arbitrarily. 
	
	Using the relation \reef{FF}, one can write the couplings \reef{S1F} in terms of dilaton  and the generalized frame. Then using the  dimensional reduction \reef{genmet1}, one can reduce the $2D$-dimensional action \reef{S1F} to the $D$-dimensional action, \eg the reduction of   $c_1$-term   is
	\beqa
\bS^f_1&=&-\frac{2 }{\kappa^2 } \int dx e^{-2\Phi}\sqrt{-G}\Big(16 c_1  D\Phi_{a} D\Phi^{a} D\Phi_{b} D\Phi^{b} + 32 c_1  De^{abi } D\Phi_{b} D\Phi_{c} 
D\Phi^{c} e_{ai }\nn\\&& - 32 c_1  De^{abi } D\Phi_{a} D
\Phi_{c} D\Phi^{c} e_{bi } + 16 c_1  De^{abi } 
De^{cdj} D\Phi_{b} D\Phi_{d} e_{ai } e_{cj} \nn\\&&- 16 
c_1  De^{abi } De_{b}{}^{cj} D\Phi_{d} D\Phi^{d} 
e_{ai } e_{cj} + 8 c_1  De^{abi } 
De^{c}{}_{b}{}^{j} D\Phi_{d} D\Phi^{d} e_{ai } e_{cj} \nn\\&&- 
32 c_1  De^{abi } De^{cdj} D\Phi_{a} D\Phi_{d} 
e_{bi } e_{cj} + 8 c_1  De_{a}{}^{cj} De^{abi } D
\Phi_{d} D\Phi^{d} e_{bi } e_{cj} \nn\\&& + 16 c_1  
De^{abi } De^{cdj} D\Phi_{a} D\Phi_{c} e_{bi } e_{dj}
- 16 c_1  De^{abi } De_{b}{}^{cj} De^{dek} D\Phi_{e} e_{ai } e_{cj} e_{dk}\nn\\&&  + 8 c_1  De^{abi } 
De^{c}{}_{b}{}^{j} De^{dek} D\Phi_{e} e_{ai } e_{cj} 
e_{dk} + 8 c_1  De_{a}{}^{cj} De^{abi } De^{dek} 
D\Phi_{e} e_{bi } e_{cj} e_{dk} \nn\\&& + 16 c_1  
De^{abi } De_{b}{}^{cj} De^{dek} D\Phi_{d} e_{ai } 
e_{cj} e_{ek} - 8 c_1  De^{abi } 
De^{c}{}_{b}{}^{j} De^{dek} D\Phi_{d} e_{ai } e_{cj} 
e_{ek}\nn\\&&  - 8 c_1  De_{a}{}^{cj} De^{abi } De^{dek} 
D\Phi_{d} e_{bi } e_{cj} e_{ek} + 4 c_1  
De^{abi } De_{b}{}^{cj} De^{dek} De_{e}{}^{fl} e_{ai } 
e_{cj} e_{dk} e_{fl}\nn\\&&  - 4 c_1  De^{abi } 
De_{b}{}^{cj} De^{dek} De^{f}{}_{e}{}^{l} e_{ai } 
e_{cj} e_{dk} e_{fl} + c_1  De^{abi } 
De^{c}{}_{b}{}^{j} De^{dek} De^{f}{}_{e}{}^{l} e_{ai } 
e_{cj} e_{dk} e_{fl}\nn\\&&  - 4 c_1  De_{a}{}^{cj} 
De^{abi } De^{dek} De_{e}{}^{fl} e_{bi } e_{cj} 
e_{dk} e_{fl} + 2 c_1  De_{a}{}^{cj} De^{abi } 
De^{dek} De^{f}{}_{e}{}^{l} e_{bi } e_{cj} e_{dk} 
e_{fl}\nn\\&&  + c_1  De_{a}{}^{cj} De^{abi } 
De_{d}{}^{fl} De^{dek} e_{bi } e_{cj} e_{ek} e_{fl}+\cdots\Big)\labell{S1F2}
\eeqa
where $De_{ab}{}^{i}\equiv\prt_a e_b{}^i$ and $D\Phi_{a}\equiv\prt_a\Phi$.   The above action is manifestly invariant under T-duality as its parents \reef{S1F} are invariant under the $O(D,D)$ transformations, however, it is not invariant under the usual $D$-dimensional gauge transformations for arbitrary parameters. We are going to find these parameters by comparing them with a   $D$-dimensional action at order $\alpha'$ which is not invariant under the T-duality, but is invariant under the conventional gauge transformations, \ie the standard coordinate transformations,  the  B-field gauge transformations and the non-standard Lorentz transformation of the B-field which is required for anomaly cancellations.

The most general $D$-dimensional action which is invariant under the conventional gauge transformations has   four class of terms. One class contains terms that are zero by Bianchi identities, one class contains terms that are total derivative terms, one class contains terms that are reproduced by the field redefinitions at order $\alpha'$ of the fields in \reef{S0bf}, and the last class contains all other terms in which we are interested. One may choose the couplings to be  \cite{Metsaev:1987zx} 
\beqa
 \bS_1^c&=&\frac{-2}{\kappa^2}\alpha'\int d^{d+1}x e^{-2\Phi}\sqrt{-G}\Big( b_1  R_{a b c d} R^{a b c d} +b_2 R_{a b c d} H^{a be}H^{ c d}{}_{e}\nn\\
&&+b_3H_{fgh}H^{f}{}_{a}{}^{b}H^{g}{}_{b}{}^{c}H^{h}{}_{c}{}^{a}+b_4H_{f}{}^{ab}H_{gab}H^{fch}H^{g}{}_{ch}+b_5(H^2)^2+b_6H_{acd}H_b{}^{cd}\prt^a\Phi\prt^b\Phi\nn\\&&
+b_7H^2\prt_a\Phi\prt^a\Phi+b_8(\prt_a\Phi\prt^a\Phi)^2+d_1 H^{abc}\Omega_{abc}\Big)\labell{S1b}
\eeqa
 where  $b_1,b_2,\cdots,b_8 $   are eight parameters. The field redefinitions freedom allows us to choose the eight arbitrary couplings in \reef{S1b} in many different schemes. The above is one particular scheme.  The last term is zero for the bosonic string theory because the B-field gauge transformation is the standard   transformation, \ie $B_{ab}\rightarrow B_{ab}+\prt_{[a}\lambda_{b]}$. This term is non-zero for the heterotic string theory and its coefficient is $d_1=-a_1/6$ in this case. This term is a result of  non-standard gauge transformation of $B$-field which is 
\beqa
B_{ab}\rightarrow B_{ab}+\prt_{[a}\lambda_{b]}+\alpha' \prt_{[a}\Lambda_i{}^j\omega_{b]j}{}^i
\eeqa
where $\Lambda_i{}^j$ is the matrix of the   Lorentz transformations and $\omega_{bi}{}^j$ is spin connection\footnote{At order $\alpha'$, there is a scheme   in which the appearance of the B-field   in the effective action  simplifies through extending the spin connection to the spin connection with torsion \cite{Bergshoeff:1989de}. Since we are working with the most general covariant action \reef{S1b}, we do not assume such simplification. }.
The  Chern-Simons three-form $\Omega$ which is defined as
\beqa
\Omega_{abc}&=&\omega_{[a i}{}^j\prt_b\omega_{c]j}{}^i+\frac{2}{3}\omega_{[ai}{}^j\omega_{bj}{}^k\omega_{c]k}{}^i\,\,;\,\,\,\omega_{ai}{}^j=\prt_a e_b{}^j e^b{}_i-\Gamma_{ab}{}^c e_c{}^j e^b{}_i
\eeqa
   makes $H_{abc}+\alpha' \Omega_{abc}$ to be invariant under the Lorentz transformations, \ie $H_{abc}+\alpha' \Omega_{abc}\rightarrow  H_{abc}+\alpha' \Omega_{abc}$. The  action \reef{S1b}   is not invariant under T-duality for arbitrary  parameters   $b_1,b_2,\cdots,b_8 $. However, one expects for some specific values for these parameters,  the couplings become invariant under T-duality. We are going to find these parameters by comparing them with the manifestly T-duality invariant action \reef{S1F2}. To compare the two actions, one has to rewrite the couplings in \reef{S1b} in terms of frame $e_a{}^i$ and $B$-field. If one compares the above $D$-dimensional action with the reduced action \reef{S1F2}, one would find zero value for all parameters. So the actions are not the same. Let us check  their Lagrangians. So we have to include some total derivative terms.

To construct all $D$-dimensional total derivative terms in terms of frame $e_a{}^i$, $B$-field and dilaton $\Phi$, we have to construct all contractions  of the following tensors with metric $G^{ab}$ to produce the current $  \bI^a$: 
\beqa
&& \partial_{a}\partial_{b}\partial_{c}B_{de} +  
\partial_{a}\partial_{b}\partial_{c}e_{d}{}^{e} +  
\partial_{a}\partial_{b}\partial_{c}\Phi +  
\partial_{a}\partial_{b}\Phi \partial_{c}e_{d}{}^{e} + 
\partial_{a}\partial_{b}\Phi \partial_{c}\Phi +  
\partial_{a}\Phi \partial_{b}\Phi \partial_{c}\Phi\nn\\&& +  
\partial_{a}B_{bc} \partial_{d}\partial_{e}B_{fg} +  
\partial_{a}B_{bc} \partial_{d}\partial_{e}e_{f}{}^{g} +  
\partial_{a}B_{bc} \partial_{d}\partial_{e}\Phi +  
\partial_{a}\partial_{b}B_{cd} \partial_{e}e_{f}{}^{g} +  
\partial_{a}\partial_{b}e_{c}{}^{d} \partial_{e}e_{f}{}^{g}\nn\\&& + 
 \partial_{a}\partial_{b}B_{cd} \partial_{e}\Phi +   
\partial_{a}\partial_{b}e_{c}{}^{d} \partial_{e}\Phi +   
\partial_{a}B_{bc} \partial_{d}\Phi \partial_{e}\Phi +  
\partial_{a}e_{b}{}^{c} \partial_{d}\Phi \partial_{e}\Phi +  
\partial_{a}B_{bc} \partial_{d}B_{ef} \partial_{g}B_{hhh}\nn\\&& +   
\partial_{a}B_{bc} \partial_{d}B_{ef} \partial_{g}e_{h}{}^{hh} + 
 \partial_{a}B_{bc} \partial_{d}e_{e}{}^{f} 
\partial_{g}e_{h}{}^{hh} +  \partial_{a}e_{b}{}^{c} 
\partial_{d}e_{e}{}^{f} \partial_{g}e_{h}{}^{hh} +  
\partial_{a}B_{bc} \partial_{d}B_{ef} \partial_{g}\Phi \nn\\&&+  
\partial_{a}B_{bc} \partial_{d}e_{e}{}^{f} \partial_{g}\Phi + 
 \partial_{a}e_{b}{}^{c} \partial_{d}e_{e}{}^{f} 
\partial_{g}\Phi\labell{cur}
\eeqa
where $\prt_a e_b{}^c\equiv e^c{}_i\prt_a e_b{}^i$, $\prt_a\prt_d e_b{}^c\equiv e^c{}_i\prt_a\prt_d e_b{}^i$  and $\prt_a\prt_d \prt_f e_b{}^c\equiv e^c{}_i\prt_a\prt_d \prt_f e_b{}^i$.     Then the following expression produces all  total derivative terms:
\beqa
\bJ&=&\frac{-2}{\kappa^2}\alpha'\int dx\prt_a(e^{-2\Phi}\sqrt{-G} \bI^a)
\eeqa
One may extend the   list of currents in \reef{cur}  by  including terms $\prt_a e_b{}^i\prt_c e_{di}$ and $\prt_a\prt_b e_c{}^i\prt_d e_{ei}$ as well, however, they do not produce independent total derivative terms in $\bJ$. We have examined the equality $\bS_1^f=\bS_1^c+\bJ$ and again found zero result for all parameters. This is an indication of the observation made in \cite{Meissner:1996sa} that   fields in the conventional $D$-dimensional action \reef{S1b} are not the same as the  fields defined in the reduction of the $2D$-dimensional fields. In particular, the fields $B, e_a{}^i$ in \reef{genmet1}  are not the same as the dynamical $B$-field and frame used in \reef{S1b}. So we have to use field redefinitions on the $D$-dimensional fields in \reef{S0bf}, \reef{S1b} and then compare them with \reef{S1F2}.

The variation of action \reef{S0bf}, for $a_1=1$, under field redefinition $G_{ab}\rightarrow G_{ab}+\delta G_{ab}$, $B_{ab}\rightarrow B_{ab}+\delta B_{ab}$ and $\Phi\rightarrow \Phi+\delta \Phi$ is 
\beqa
\delta\bS_0^c&=&-\frac{2 }{\kappa^2}\int d x e^{-2\Phi}\sqrt{-G}\Big[\Big(\frac{1}{2}R +2\nabla_c\nabla^c\Phi  -2\nabla_c\Phi\nabla^c\Phi -\frac{1}{24}H^2\Big)(G^{ab}\delta G_{ab}-4\delta \Phi)\nn\\&&-\Big(R^{ab}+2\nabla^b\nabla^a\Phi-\frac{1}{4}H^{acd}H^b{}_{cd}\Big)\delta G_{ab}\labell{per}+\frac{1}{2}(\nabla_cH^{cab}-2\nabla_c\Phi H^{cab}) \delta B_{ab}\Big]
\eeqa
where we have also ignored some covariant total derivative terms. $\delta G_{ab}$, $\delta\Phi$ and $\delta B_{ab}$ can be constructed  by all contractions of the following tensors with metric $G^{ab}$:
\beqa
&& \partial_{a}\partial_{b}B_{cd} +  
\partial_{a}\partial_{b}e_{c}{}^{d} +  
\partial_{a}\partial_{b}\Phi +  \partial_{a}\Phi 
\partial_{b}\Phi +  \partial_{a}B_{bc} \partial_{d}B_{ef}\nn\\&& + 
\partial_{a}B_{bc} \partial_{d}e_{e}{}^{f} + 
\partial_{a}e_{b}{}^{c} \partial_{d}e_{e}{}^{f} +   
\partial_{a}B_{bc} \partial_{d}\Phi +   
\partial_{a}e_{b}{}^{c} \partial_{d}\Phi
\eeqa
We then examine the following equality:
\beqa
 \bS_1^f&=&\bS_1^c+\bJ+\delta \bS_0^c
\eeqa
It produces  many algebraic equations for the parameters with  non-zero result for them. 

The most important part  of the result is that they fix uniquely all eight parameters in  the $D$-dimensional action \reef{S1b} in terms of $b_1$, \ie    
\beqa
 \bS_1^c&=&\frac{-2 }{\kappa^2}\alpha'\int d^{d+1}x e^{-2\Phi}\sqrt{-G}\Big[b_1\Big(   R_{a b c d} R^{a b c d} -\frac{1}{2} R_{a b c d} H^{a be}H^{ c d}{}_{e}\nn\\
&&+\frac{1}{24}H_{fhg}H^{f}{}_{a}{}^{b}H^{h}{}_{b}{}^{c}H^{g}{}_{c}{}^{a}-\frac{1}{8}H_{f}{}^{ab}H_{hab}H^{fcg}H^{h}{}_{cg}\Big) +d_1H^{abc}\Omega_{abc}\Big]\labell{S1bf}
\eeqa
 In the scheme \reef{S1b}, the effective action \reef{S1bf} then has no derivative of dilaton. Up to the overall factor $b_1$, the above couplings are   the standard effective action of the bosonic   and heterotic string theories   which has been found in \cite{Metsaev:1987zx} by the S-matrix calculations.  This action now is invariant under T-duality. Since we have used field redefinitions to relate  the manifestly T-duality invariant action \reef{S1F2} to the manifestly gauge invariant action \reef{S1b}, the T-duality transformations rules in above action would be  the Buscher rules plus their $\alpha'$-corrections.  They have be found directly by using the T-duality approach \cite{Bergshoeff:1995cg,Kaloper:1997ux}. 

The parameters in the $2D$-dimensional action \reef{S1F}, however,  are not fixed  uniquely in terms of $b_1,\, d_1$. There remain many   parameters in \reef{S1F} to be arbitrary. They reflect the four classes  of terms  in \reef{S1F} that we did not removed them from the list of independent  couplings in \reef{S1F}. Choosing different values for the arbitrary parameters correspond to different scheme in which the $2D$-dimensional action can be written. One particular choice for the parameters should reproduce the $2D$-dimensional couplings found in \cite{Baron:2017dvb}  which include derivatives of the generalized dilaton and frame. However, as in the $D$-dimensional action \reef{S1bf}, we choose a scheme in which the $2D$-dimensional dilaton appears only as an overall factor. Since the derivative of the dilaton appears in the flux $\cF_A$, we choose the arbitrary parameters to have no terms in \reef{S1F} which has $\cF_A$ or its derivatives. Using this constraint, there are still   some residual parameters in this scheme. We further constrain the scheme to have   no terms with structure $\prt_A\prt_B\prt_C\cF_{DEF}$, $\prt_A\prt_B \cF_{CDE}$, nor $\prt_A\cF_{BCD}\prt_{E}\cF_{EFH}$. Then the effective action in this scheme for the bosonic string theory becomes 
\beqa
\bS_1^{fe}&=&\frac{-2b_1}{\kappa^2\tV}\alpha' \int dx d\tilde{x} e^{-2d}\Big(-  \frac{1}{8} 
\cH^{A B} \cH^{C D} \cH^{E 
F} \cH^{G H} \cH^{I J} 
\cH^{K L} \cF_{A C E} 
\cF_{B D G} \cF_{F 
I K} \cF_{H J L}  \nn\\&&+
\frac{1}{4} \cH^{A B} \cH^{C D} 
\cF_{A}{}^{E F} \cF_{B}{}^{G H} \cF_{C E 
[F} \cF_{G]D   H} + \frac{1}{4} \cH^{A B} 
\cH^{C D} \cH^{E F} \cH^{G 
H} \cF_{A C E} \cF_{B D G} \cF_{F}{}^{I J} \cF_{H I J} \nn\\&& + \cH^{A B} \cH^{C D} \cF_{A C}{}^{E} \cF_{B}{}^{F G} 
\cF_{D F}{}^{H} \cF_{E G H} + \frac{1}{2} \cH^{A B} 
\cH^{C D} \cF_{A C}{}^{E} \cF_{B E}{}^{F} \cF_{D}{}^{G H} \cF_{F G 
H} \nn\\&& + \frac{3}{8} \cH^{A B} \cH^{C 
D} \cH^{E F} \cH^{G H} 
\cF_{A C}{}^{I} \cF_{B 
E}{}^{J} \cF_{D G J} 
\cF_{F H I} \nn\\&&-  \frac{1}{2} 
\cH^{A B} \cH^{C D} \cH^{E 
F} \cH^{G H} \cF_{A C 
E} \cF_{B G}{}^{I} \cF_{D H}{}^{J} \cF_{F I J} \labell{S1Ff}\\&& + \frac{1}{24} \cH^{A B} \cH^{C 
D} \cH^{E F} \cH^{G H} \cH^{
I J} \cH^{K L} \cF_{A 
C E} \cF_{B G I} 
\cF_{D H K} \cF_{F 
J L}  \nn\\&&-2 
\cH^{A B} \cH^{C D} \cF_{A 
C}{}^{E} \cF_{B}{}^{F 
G}\prt_{[F}\cF_{E]  D   G} + 
  \cH^{A B} \cH^{C D} 
\cH^{E F} \cH^{G H} \cF_{A C E} \cF_{B G}{}^{
I}\prt_{(H}\cF_{I) D F  } +\cdots \Big)\nn
\eeqa
where dots represent terms that are zero under the strong  constraint. Our notation for anti-symmetrization is $L_{[A}P_{B]}=(L_AP_B-L_BP_A)/2$, similarly for symmetrization. This action now must be  invariant under double-Lorentz transformations. Since we have used field redefinitions to relate  the manifestly T-duality invariant action \reef{S1F2} to the manifestly Lorentz invariant action \reef{S1b}, the double-Lorentz transformations    in above action would be  the standard Lorentz transformations  plus their $\alpha'$-corrections. The above action is even under $B\rightarrow -B$. It should be the same as the even part of the action has been found in \cite{Baron:2017dvb} up to the terms that are zero under the strong  constraint and up to $2D$-dimensional field redefinitions.

The $2D$-dimensional effective action of the heterotic string theory at order $\alpha'$ contains the above action plus the following terms which are odd under $B\rightarrow -B$:
\beqa
\bS_1^{fo}&=& \frac{-2d_1}{ \kappa^2\tV}\alpha' \int dx d\tilde{x} e^{-2d}\Big(\frac{1}{8}   \cH^{A B} \cH^{C D} 
\cH^{E F} \cF_{A C}{}^{G} \cF_{B D}{}^{ I} \cF_{E G}{}^{H} \cF_{F 
 I H}\nn\\&& -  \frac{1}{8}   \cH^{A B} 
\cH^{C D} \cH^{E F} \cH^{G 
 I} \cH^{H J} \cF_{A C 
E} \cF_{B G H} \cF_{D  I}{}^{K} \cF_{F J K} \nn\\&&- 
 \frac{1}{12}   \cH^{A B} \cH^{C D} \cH^{E F} \cF_{A 
C}{}^{G} \cF_{B E}{}^{ I} 
\cF_{D F}{}^{H} \cF_{G  I H} + \frac{1}{24}   \cH^{A B} \cH^{
C D} \cH^{E F} \cF_{A C E} \cF_{B D}{}^{G} \cF_{F}{}^{ I H} 
\cF_{G  I H}\nn\\&& + \frac{1}{4}   \cH^{A B} \cH^{C D} \cH^{E F} 
\cF_{A C}{}^{G} \cF_{E 
G}{}^{ H} \prt_B\cF_{  D F 
 H} - \frac{1}{3}   \cH^{A B} \cH^{C 
D} \cH^{E F} \cF_{A 
C E} \cF_{B}{}^{G  H} 
\prt_{[D}\cF_{ G] F    H} \nn\\&&-  
\frac{1}{4}   \cH^{A B} \cH^{C D} 
\cH^{E F} \cH^{G  I} \cH^{H 
J} \cF_{A C E} \cF_{B G H} \prt_D\cF_{  F 
 I J} -  \frac{1}{8}   \cH^{A B} 
\cH^{C D} \cH^{E F} \cF_{A C}{}^{G} \cF_{E G}{}^{ H} \prt_F\cF_{  B D  H}\nn\\&& 
 -  \frac{3}{8}   
\cH^{A B} \cH^{C D} \cH^{E 
F} \cF_{A C}{}^{G} \cF_{E G}{}^{ H}\prt_H \cF_{   B 
D F}+\cdots \Big)\labell{S1F3}
\eeqa
where dots represent terms that are zero under the strong  constraint. While the action \reef{S1Ff} has even number of $\cH^{AB}$, the above action has odd number of  $\cH^{AB}$. The above action  should be the same as the odd part of the action has been found in \cite{Baron:2017dvb} up to the terms that are zero under the strong  constraint and up to $2D$-dimensional field redefinitions.

The algebraic equations    fix also the non-covariant field redefinitions and total derivative terms required to relate  the two $D$-dimensional actions. The   total derivative terms in the bosonic theory  have    structures $\prt\prt e\prt\prt e$, $\prt\prt e \prt e\prt e$, $\prt\prt B\prt B\prt e$ , $\prt B\prt B\prt\prt e$, $\prt B\prt B\prt e\prt e$, $\prt\prt\Phi\prt e\prt e$, $\prt\prt\Phi\prt\Phi\prt e$, $\prt\Phi\prt e\prt\prt e$, $\prt\Phi\prt\Phi\prt e\prt e$, $\prt\Phi\prt e\prt e\prt e$ and $\prt\Phi\prt e\prt B\prt B$. They indicate that the current $\bI^a$ contain terms with structure $\prt\prt e\prt e$, $\prt e\prt e\prt e$, $\prt\Phi\prt e\prt e$ and $\prt e\prt B\prt B$. The reason that there is no current with structure $\prt\prt\Phi\prt e$, $\prt\prt\Phi\prt\Phi$ , $\prt\prt B\prt B$ is that they produce terms in $\bJ$ with structures $\prt\prt\Phi\prt\prt\Phi$, $\prt\prt\Phi\prt\prt e$, $\prt\prt B\prt\prt B$ which are not in the $D$-dimensional action \reef{S1b}.  There are too many total derivative terms to be able to write them explicitly here.  The field redefinitions for the bosonic theory however are
\beqa
\delta G_{ab}^B&=& b_1\Big(2 De_{a}{}^{ci} De_{bci} -  4 De_{(b}{}^{ci} De_{ca)i} + 
2 De_{cbi} De^{c}{}_{a}{}^{i} + 8 De_{(b}{}^{ci} D\Phi _{c} e_{a)i} - 8 De^{c}{}_{(b}{}^{i} D\Phi _{c} e_{a)i}\nn\\&& - 
4 DDe_{(b}{}^{c}{}_{ci} e_{a)}{}^{i} + 4 DDe^{c}{}_{c(bi} 
e_{a)}{}^{i} - 4 De_{cd}{}^{j} De^{cdi} e_{ai} e_{bj} + 
4 De^{cdi} De_{dc}{}^{j} e_{ai} e_{bj}\nn\\&& + 4 
De_{(b}{}^{ci} De^{d}{}_{d}{}^{j} e_{a)i} e_{cj}  - 4 
De^{c}{}_{(a}{}^{i} De^{d}{}_{d}{}^{j} e_{b)i} e_{cj} + 2 
De_{(b}{}^{ci} De_{c}{}^{dj} e_{a)j} e_{di} - 6 
De_{c}{}^{dj} De^{c}{}_{(b}{}^{i} e_{a)j} e_{di}\nn\\&& - 2 
De_{(a}{}^{ci} De^{d}{}_{c}{}^{j} e_{b)j} e_{di} + 6 
De^{c}{}_{(a}{}^{i} De^{d}{}_{c}{}^{j} e_{b)j} e_{di} + 2 
De_{a}{}^{ci} De_{b}{}^{dj} e_{cj} e_{di} - 4 
De_{(a}{}^{ci} De^{d}{}_{b)}{}^{j} e_{cj} e_{di} \nn\\&&+ 2 
De^{c}{}_{a}{}^{i} De^{d}{}_{b}{}^{j} e_{cj} e_{di} - 4 De_{(a}{}^{ci} De_{c}{}^{dj} e_{b)i} e_{dj} +4 
De_{c}{}^{dj} De^{c}{}_{(a}{}^{i} e_{b)i} e_{dj} + 4 
De_{(a}{}^{ci} De^{d}{}_{c}{}^{j} e_{b)i} e_{dj}\nn\\&& - 4
De^{c}{}_{(a}{}^{i} De^{d}{}_{c}{}^{j} e_{b)i} e_{dj}\Big)\nn\\
\delta\Phi^B&=&\frac{1}{4}G^{ab}\delta G_{ab}^B\nn\\
\delta B_{ab}^B&=&b_1H^{cd}{}_{[a}(\omega_{b]ij}e_c{}^ie_d{}^j+e_{b]i}De_{cd}{}^i)\labell{f1}
\eeqa
where $DDe_{abc}{}^i\equiv \prt_a\prt_b e_c{}^i$,  $De_{ab}{}^{i}\equiv\prt_a e_b{}^i$ and $D\Phi_{a}\equiv\prt_a\Phi$. The curved indices are raised with metric $G^{ab}$ and flat indices are lowered by $\eta_{ij}$. The above $\alpha'$-corrections must be  added to the metric, dilaton and the B-field in the reductions of $2D$-dilaton, \ie $e^{-2d}=e^{-2\Phi}\sqrt{-G}$ and the generalized metric \reef{genmet} in order to make them  identical to the corresponding covariant fields in the $D$-dimensional action \reef{S1b}. The resulting new $2D$-dilaton $\tilde{d}$ and generalized metric $\tilde{\cH}$, then should transform covariantly as \reef{gcov}. The above field redefinition for dilaton, however,  is such that $\delta d=\delta \Phi-\frac{1}{4}G^{ab}\delta G_{ab}=0$. So the generalized dilaton remains invariant under the field redefinitions at order $\alpha'$. Using the  $\alpha'$-corrections to the generalized metric, one may find the  $\alpha'$-corrections to the generalized  diffeomorphisms for the fields that are transformed by the Buscher rules under T-duality.  Alternatively, one may add the above $\alpha'$-corrections, with a minus sign, to the $D$-dimensional covariant fields $G_{ab},\Phi, B_{ab}$ in order to make them identical to the corresponding $2D$-dimensional fields $e^{-2d}=e^{-2\Phi}\sqrt{-G}$ and   \reef{genmet} which are transformed  by Buscher rules  under T-duality\cite{Aldazabal:2013sca}. In this way one may find the $\alpha'$-corrections to the Buscher rules for the  $D$-dimensional covariant fields.

The   total derivative terms in the heterotic theory  have  the same   structures as in the bosonic theory as well as the structures $\prt\prt e\prt\prt B$, $\prt\prt e \prt e\prt B$, $\prt e\prt e\prt \prt B$ , $\prt e\prt e\prt\Phi\prt B$, $\prt e\prt \Phi\prt  \prt B$, $\prt\prt e\prt \Phi\prt B$, $\prt\Phi\prt\Phi\prt e\prt B$ and $\prt e\prt \prt\Phi\prt B$. They indicate that the current $\bI^a$ contain terms with structure $\prt\prt e\prt B$, $\prt e\prt e\prt B$,   and $\prt e\prt \Phi\prt B$.   The field redefinitions for the heterotic theory are
\beqa
\delta G_{ab}^H&=&\delta G_{ab}^B+d_1H^{cd}{}_{(a} e_{b)i}De_{cd}{}^i\nn\\
\delta \Phi^H&=&\frac{1}{4} G^{ab}\delta G_{ab}^H\nn\\
\delta B_{ab}^H&=&\delta B_{ab}^B+d_1\Big(De_{[ab]}{}^{i} De^{c}{}_{ci} + DDe_{[a}{}^{c}{}_{ci} 
e_{b]}{}^{i} - 4 De_{[ab]}{}^{i} D\Phi^{c} e_{ci} + 
DDe_{[acb]i} e^{ci}\nn\\&&  + De_{c}{}^{dj} De^{c}{}_{[b}{}^{i} 
e_{a]j} e_{di}- 2 De_{[a}{}^{ci} De^{d}{}_{c}{}^{j} 
e_{b]j} e_{di} + De^{c}{}_{[a}{}^{i} De^{d}{}_{c}{}^{j} 
e_{b]j} e_{di} \nn\\&&+ 2 De_{[ba]}{}^{i} De^{cdj} e_{cj} 
e_{di} -  De_{[a}{}^{ci} De^{d}{}_{b]}{}^{j} e_{cj} 
e_{di} + 2 De_{[ab]}{}^{i} De^{cdj} e_{ci} e_{dj}\Big)\labell{f2}
\eeqa
Here also the field redefinition for the generalized dilaton $\delta d$ is zero. While the   field redefinition terms in the bosonic theory do not change the symmetry under $B\rightarrow -B$, the field redefinition terms corresponding to the Chern-Simons term in the heterotic theory change the symmetry of fields  under  $B\rightarrow -B$. This is as expected because the field redefinitions should produce odd terms under $B\rightarrow -B$ from the even terms in \reef{per}. It is interesting to note that in both heterotic and bosonic theories the total derivative term are zero when the $D$-dimensional frame $e_a{}^i$ is constant. Moreover, the T-duality covariant fields and the gauge covariant fields are identical when  $e_a{}^i$ is constant, \ie $\delta G=\delta \Phi=\delta B=0$.

We have found the T-duality invariant  2$D$-dimensional actions \reef{S1Ff} and  \reef{S1F3} by constraining them to be invariant under the generalized diffeomorphism and by constraining them to be invariant under the  $D$-dimensional gauge transformations after using appropriate field redefinitions \reef{f1} and \reef{f2} on the leading order action \reef{ODD1}. These actions must be invariant under  the double-Lorentz transformations as well. However, the transformations  are the standard double-Lorentz transformations pulse their corresponding $\alpha'$-corrections \cite{Marques:2015vua,Baron:2017dvb}. Since we have used the $2D$-dimensional field redefinitions freedom to write the $\alpha'$-order actions \reef{S1Ff} and  \reef{S1F3} in a scheme in which the  dilaton appears only as an overall factor,  the field redefinitions \reef{f1} and \reef{f2} and the corresponding  $\alpha'$-corrections to the double-Lorentz transformations are then fixed uniquely. If we have chosen a different scheme for the $2D$-dimensional actions, then the field redefinitions  \reef{f1} and \reef{f2} would be different. The form of the corrections to the double-Lorentz transformations might  be different from those have been found in \cite{Marques:2015vua,Baron:2017dvb}, because the $2D$-dimensional actions in \cite{Marques:2015vua,Baron:2017dvb} is written in a different scheme in which the derivative of dilaton also appears in the actions. Unlike our approach that    the $2D$-dimensional action is  fixed first,   in \cite{Marques:2015vua,Baron:2017dvb} the authors first fix the $\alpha'$-corrections to the  double-Lorentz transformations and then find the corresponding $2D$-dimensional actions.   It would be interesting to find the $\alpha'$-corrections to the  double-Lorentz transformations corresponding to the actions \reef{S1Ff} and  \reef{S1F3} and then compare them with the   transformations found in   \cite{Marques:2015vua,Baron:2017dvb} under the  appropriate field redefinitions that change the actions \reef{S1Ff} and  \reef{S1F3} to the corresponding actions in  \cite{Marques:2015vua,Baron:2017dvb}.

Using the  generalized frame and dilaton as the dynamical fields, we have found the $2D$-dimensional actions \reef{S1Ff} and \reef{S1F3}. One may wish to use the generalized metric and dilaton as the dynamical fields to find the $2D$-dimensional action at order $\alpha'$, \ie extension of   the action \reef{ODD1} to the order $\alpha'$.   We have done this calculation. We have found that for the case that B-field is zero there are non-zero $D$-dimensional and $2D$-dimensional actions, however, in the presence of the B-field one would find no effective action at order $\alpha'$. In fact, when we write the couplings \reef{S1b}  in terms of B-field and metric, for instance, the second coupling in \reef{S1b} produces, among other things,  the following terms: 
 \beqa
-2   G^{fg} \partial_{a}G_{dg} \partial^{a}B^{bc} 
\partial_{b}B^{de} \partial_{c}G_{ef} +   G^{fg} 
\partial_{a}B^{de} \partial^{a}B^{bc} \partial_{b}G_{dg} 
\partial_{c}G_{ef}
\eeqa
None of them is reproduced by any $2D$-couplings, any non-covariant field redefinition or any total derivative term  at order $\alpha'$. This confirms the observation that there is no manifestly background independent and duality invariant formulation of the bosonic theory at order $\alpha'$ in terms of the generalized metric, however, there is such formulation in terms of generalized frame  \cite{Hohm:2016lge}.  So one expects the convenient frame-work for studying the higher derivative couplings in the $D$-dimensional string theory and in DFT is the frame-like formulation of DFT.   

The gravity and dilaton couplings in the effective actions of string theories at orders $\alpha'^2$ and $\alpha'^3$ are known in the literature. Using the T-duality approach, it has been shown in  \cite{Razaghian:2017okr,Razaghian:2018svg} that they are  invariant under the T-duality transformations which are the Buscher rules and their $\alpha'$-corrections. However, the B-field couplings at these orders are not known in the literature. It would be interesting to use the method in this paper to find these couplings as well as their corresponding $2D$-dimensional actions. The B-field couplings at order $\alpha'^2$ for Hohm-Siegel-Zwiebach double field theory \cite{Hohm:2013jaa} have been found in \cite{Lescano:2016grn}.

{\bf Acknowledgments}:   This work is supported by Ferdowsi University of Mashhad under grant  2/46948(1397/04/05).

\end{document}